\documentclass[conference]{IEEEtran}
\IEEEoverridecommandlockouts
\usepackage[style=numeric]{biblatex}
\usepackage{url}
\usepackage{siunitx}
\usepackage{booktabs}
\usepackage{amsmath}
\usepackage{graphicx}
\usepackage{amssymb}
\usepackage{amsfonts}
\usepackage{algorithm}
\usepackage{algpseudocode}
\usepackage{fancyhdr}
\usepackage{caption}
\usepackage{pgf}
\usepackage{pgfplots}
\pgfplotsset{width=10cm,compat=1.9}
\usepackage{tikz}
\usepackage{tikz-cd}
\usepackage{subcaption}
\usepackage{tabularx}
\usepackage{varwidth}
\usepackage[nameinlink,noabbrev,capitalize]{cleveref}
\pgfkeys{/pgf/number format/.cd,1000 sep={\,}}
\usepgfplotslibrary{groupplots}
\usetikzlibrary{shapes.geometric, arrows, positioning, calc, fit}
\usepackage{bbding}

\newcommand{\nint}[1]{\left\lfloor #1 \right\rceil}

\algdef{SE}[FUNCTION]{Function}{EndFunction}%
   [2]{\algorithmicfunction\ \textproc{#1}\ifthenelse{\equal{#2}{}}{}{(#2)}}%
   {\algorithmicend\ \algorithmicfunction}%

\algnewcommand\algorithmicforeach{\textbf{for each}}
\algdef{S}[FOR]{ForEach}[1]{\algorithmicforeach\ #1\ \algorithmicdo}

\def\BibTeX{{\rm B\kern-.05em{\sc i\kern-.025em b}\kern-.08em
    T\kern-.1667em\lower.7ex\hbox{E}\kern-.125emX}}

\begin{document}

\title{De-DSI: Decentralised Differentiable Search Index}

\author{\IEEEauthorblockN{Petru Neague}
\IEEEauthorblockA{\textit{Delft University of Technology} \\
Delft, Netherlands \\
p.m.neague@tudelft.nl}
\and
\IEEEauthorblockN{Marcel Gregoriadis}
\IEEEauthorblockA{\textit{Delft University of Technology} \\
Delft, Netherlands \\
m.gregoriadis@tudelft.nl}
\and
\IEEEauthorblockN{Johan Pouwelse}
\IEEEauthorblockA{\textit{Delft University of Technology} \\
Delft, Netherlands \\
j.a.pouwelse@tudelft.nl}

}

\maketitle

\begin{abstract}

  This study introduces De-DSI, a novel framework that fuses large language models (LLMs) with genuine decentralization for information retrieval, particularly employing the differentiable search index (DSI) concept in a decentralized setting. Focused on efficiently connecting novel user queries with document identifiers without direct document access, De-DSI operates solely on query-docid pairs. To enhance scalability, an ensemble of DSI models is introduced, where the dataset is partitioned into smaller shards for individual model training. This approach not only maintains accuracy by reducing the number of data each model needs to handle but also facilitates scalability by aggregating outcomes from multiple models. This aggregation uses a beam search to identify top docids and applies a softmax function for score normalization, selecting documents with the highest scores for retrieval. The decentralized implementation demonstrates that retrieval success is comparable to centralized methods, with the added benefit of the possibility of distributing computational complexity across the network. This setup also allows for the retrieval of multimedia items through magnet links, eliminating the need for platforms or intermediaries.
\end{abstract}

 \maketitle

\section{Introduction}

The proliferation of smart devices is raising concerns over privacy due to extensive data collection. 
This data is valuable for enhancing machine learning (ML) models but raises a significant risk of personal surveillance and loss of privacy. 
While the issue concerns many types of data (e.g. geolocation, personal conversations, health data, etc.), this paper attempts to take a step towards the maintenance of privacy in the field of information retrieval.

Google's federated learning, introduced in 2016, addresses privacy concerns by allowing devices to contribute to ML model improvement without sharing local data. They only exchange model updates with one or more central servers~\cite{mcmahan2023communicationefficient}.

Gossip learning, a subset of algorithms of what later came to be called \emph{decentralised federated learning}, has been proposed in 2011 to resolve the same challenge as federated learning~\cite{ormandi2011efficient,Ormándigossiplearning}.
Gossip learning is fully decentralised and does not require a central server.
Participants communicate directly, exchange model updates, and aggregate said updates. The great advantage of gossip learning is the lack of any infrastructure, making it both robust and easier to scale.
Decentralised federated learning is now becoming a mainstream topic as studied by \cite{Mart_nez_Beltr_n_2023}. 
However, to date, Bitcoin and BitTorrent are the only examples of full decentralisation with actual broad usage.
Decentralised federated learning still remains constrained to the lab~\cite{Mart_nez_Beltr_n_2023}.
Originally, decentralisation guided the development of the Internet.
One of the Internet's defining principles is its lack of any single point of technical, political, or economic control~\cite{ietf9518Centralization}.

Transformers and generative AI tuned for search are changing the field of information retrieval.
AI-based alternatives now exist for search engine strategies such as keyword matching, BM25, IDF heuristics, 
and relevance ranking~\cite{Google2001}.
Traditional IR systems separate the steps of indexing, retrieval, and (re)ranking.
One problem with the classical paradigm is that it is difficult to optimize various components. The various modules operate mostly separately,
potentially producing suboptimal results for the architecture as a whole \cite{llmSurvey}.

The leap in AI research has provoked the emergence of retrieval systems with generative models that do not rely on an explicit index anymore \cite{metzler2021rethinking,zhou2023dynamicretriever,chen2022corpusbrain}.
Instead, the knowledge of the documents is encoded in the parameters of a pre-trained model.
The idea is that those models will be able to `understand' queries and documents,
rather than just remember and match,
and as a result, be better at retrieval.
Moreover, the model can compress more information,
allowing generative retrievers to occupy significantly less space
than traditional retrieval methods.

The first wave of this research has explored the direct extraction of knowledge,
where the model generates the answer to a question
after training on a corpus of documents
\cite{petroni2019language,petroni2020context,roberts2020much}.
Recently a new line of work has come up
that investigates the generation of document identifier strings directly from a model~\cite{tay2022transformer}.
This novel search architecture is called \textit{Differentiable Search Index (DSI)}.

DSI uses a single Transformer model to perform both indexing and retrieval.
It is co-invented by three Big Tech companies.
Meta published the initial sketches for entity retrieval in 2020~\cite{de2020autoregressive};
Google introduced generic information retrieval in early 2022~\cite{tay2022transformer};
Microsoft released \emph{DSI-QG} (DSI with query generation) improvements in late 2022~\cite{zhuang2022bridging}.

We present De-DSI, the first successful fusion of two powerful, yet largely unexplored fields within machine learning. De-DSI combines \emph{Decentralised Federated Learning} (DFL) with \emph{Differentiable Search Index (DSI)}. 
The contributions of this work are the insights that search could be powered by LLMs trained in a decentralised manner, and that those could be used to build a decentralised public search engine. Due to our academically-pure decentralisation, this experimental search engine is owned and controlled by no one entity.
We trained the Google T5 model to output document IDs in response to queries in a decentralised environment. 
By using sharding, we craft an ensemble of models which allows the indexing of up to 10 times more data than a single model, at the cost of accuracy.

\section{Problem Description}

Developing a decentralised search engine has proven to be difficult. The main problems are the huge amounts of information to index, rampant online fraud, and high expectations of users (near-perfect, near-instantaneous results). 
\textit{Differentiable Search Index (DSI)} represents a promising paradigm for information retrieval tasks using Transformers~\cite{tay2022transformer}. 
Assessing the viability of DSI within real systems with real users and enormous amounts of data is still unsolved. DSI might prove to be similar to the DHT: elegant and ineffective. It is an emerging paradigm for information retrieval, yet it still lacks decentralisation, scalability, security, privacy, practical validation, and user trails.

Numerous scientists have investigated algorithms for \emph{distributed} information retrieval~\cite{Hiemstra2012}.
However, transforming these ideas into sustainable solutions without any single point of technical, political, or economic control remains unsolved.
The demise of numerous open search engine projects contains important learnings on the challenges for long-enduring sustainable solutions.

A simple query flooding approach across a peer-to-peer overlay network was first used by Gnutella in 2001~\cite{ripeanu2001peer}.
This popular system slowly collapsed and showed the need for effective search, free-riding prevention, and anti-spam measures.
The YaCY search engine used a DHT to store the reverse word index in 2003. This DHT resulted in unsolved security issues such as spamming, poisoning, and sybil attacks~\cite{urdaneta2011survey}. 

Ultimately, it proved to be less scalable than then believed. This is caused by the mechanism to implicitly address churn and re-announce \emph{every} document daily~\cite{ipfsAminothe}.
The leading search engine for the IPFS distributed content sharing system was shut down in 2023, after seven years of operation~\cite{ipfssearchBumpRoad}.
Basic features such as relevance ranking for random files shared on IPFS proved to be difficult to realise in a fully decentralised manner. Their central website, expensive 100-node cluster, and algorithmic improvements proved to be unsustainable without continuous grant money~\cite{githubDevipfssearch}.

\section{De-DSI: Architecture and Design}

Our proof-of-concept for De-DSI is simplistic yet capable of offering effective search.
The cardinal design principles of our design are simplicity, scalability, feasibility, and deployability. 
We have access to real-world search workloads due to prior decentralised systems research which received several million user installs~\cite{pouwelse2000open,wang2005wi,pouwelse2008tribler, qwertycubeGitHubRelease}.
Our De-DSI design and experiments are devised to realistically reflect our production environment.
We aim to deploy and iteratively improve De-DSI for the coming years. 
Specifically, we plan to use it to enhance our open-source, decentralised, YouTube-like video search engine, which boasts 2.4 million unique installs as of February 2024~\cite{qwertycubeGitHubRelease}.
Our implementation of De-DSI, along with the experimental setups, is available online as open source.\footnote{\url{https://github.com/pneague/De-DSI}}

\subsection{Differentiable Search Index (DSI)}\label{sec: DSI Intro}

\begin{table}
\caption{Example of queries for a document ID. The query is the input into the DSI model and the docid is the response. The docid is constructed by the model token-by-token}
\centering
\begin{tabular}{ll}
\toprule
Query & Docid \\
\midrule
aarp spider solitaire free game & D3125778 \\
spider solitaire free game & D3125778 \\
spidersolitairefree & D3125778 \\
free spider solitaire card game & D3125778 \\
solitaire spider free & D3125778 \\
free solitaire spider games & D3125778 \\
\bottomrule
\end{tabular}
\label{table: Sample for DSI model}
\end{table}
Our De-DSI design is inspired by the original DSI work from 2022~\cite{tay2022transformer}. 
In the original DSI, a single T5-based Transformer is trained to directly map queries to document identifiers (docid), in a sequence-to-sequence fashion.
To this end, they trained the model on data from the \textit{Natural Questions (NQ)} dataset~\cite{kwiatkowski2019natural}.
Specifically, their model was trained in two phases:
initial training to associate document content with a docid, and \emph{fine-tuning} to associate each question with a matching docid. De-DSI simplifies this approach by only requiring user queries.

The \emph{structured} approach within the original DSI generates the docid token by token. This method boasts enhanced scalability compared to the use of unstructured atomic identifiers (where each document is associated to exactly one token), as it effectively narrows down the search space with each step. 
Semantically structured identifiers, where each token choice in a docid is inherently meaningful, are the most advanced form of docids for the retrieval task.

The focus of our work is the retrieval of files in decentralised networks. Such systems notoriously lack good-quality metadata (often only being given the file name). This is the reason why we chose to train the model to associate queries with the docid. 
It is also the reason why we use the \textit{naively structured} identifiers to represent our documents (i.e. the docids are represented by a sequence of randomly assigned characters with no inherent meaning).
We show a few samples of input-output pairs belonging to one document in \cref{table: Sample for DSI model}.
The DSI method allows for retrieval of multiple ranked docids through beam search. In our investigations, we find that beam search sometimes results in hallucinations. However, most of the time, its outputs yield reasonable responses. 
We adopt this method when investigating the metric of top-$k$ accuracy.
That is, we count it as a success if the expected docid is within the top-$k$ docids retrieved by the model.

\subsection{Ensemble DSI} \label{sec: Ensemble description}

The effectiveness of DSI is inherently tied to the model's size, as there is a finite limit to the amount of information a model can fit.
Since the T5-small model has fewer weights compared to models representing the state-of-the-art, scalability is bound to become a problem in systems spanning hundreds of millions of documents.
Our approach of distributing many T5-small instances in a peer-to-peer (P2P) network improves the scalability of DSI.

To this end, we propose \emph{sharding} as a means to divide the document space,
and moreover divide the information load on individual models.
Specifically, we propose splitting peers into groups.
Each group is then responsible for only one partition (one \emph{shard}) of the data.
This means they fine-tune their T5 models on all query-docid mappings that are associated with a specific subset of the existing documents.
The size of the group merely promotes robustness,
since peers within the group achieve convergence by training on the same dataset.
Yet, the peers only ever become aware of a subset of documents, and are oblivious to others.
In order to process unseen queries, therefore, 
peers must consult every other group of peers,
for the chance that their shard contains the relevant documents.

Taking into consideration the suggestions from all peer groups means that some might be valid and relevant (as they have seen similar queries),
and some will not.
Hence arises the challenge of differentiating between these results.
To this end, we make use of the logit scores attached with each produced output.
Those scores can be thought of as the model's confidence about the respective output.
More specifically, we let every peer group return the five most likely outputs given the query, using beam search.
In the case that the model has seen similar queries in its training phase,
it will be very confident about the document associated with said queries, and much less confident about other documents it knows.
Meanwhile, models which have not seen similar queries are prone to produce a random set of docids they have learnt of, each with roughly the same probability. 
However, as has been pointed out by Zhou et al.~\cite{zhou2023dynamicretriever},
these models learn on different scales, and so the scores are therefore not directly comparable.

\begin{figure}[h]
    \centering
    \tikzstyle{block} = [rectangle, draw, fill=blue!20, text centered, rounded corners]
\tikzstyle{line} = [draw, -latex]
\tikzstyle{body} = [draw, rectangle, rounded corners, minimum width=2cm, align=left, font=\scriptsize]
\tikzstyle{header} = [below=0cm of shard1.north, minimum width=2cm, minimum height=0.4cm, font=\footnotesize]

\begin{tikzpicture}[node distance=2cm, auto]

    \node[] (query) at (2,1.7) {Query};
    
    \node[body, anchor=east] (shard1) at (0,0) {
        \\ \\
        $\bullet$ Doc A1 (0.6) \\
        $\bullet$ Doc A2 (0.15) \\
        $\bullet$ Doc A3 (0.1) \\
        $\bullet$ Doc A4 (0.09) \\
        $\bullet$ Doc A5 (0.06)
      };
    \node[header, below=0cm of shard1.north] {\textbf{Shard A}};
    \draw[] ([yshift=-0.4cm]shard1.north west) -- ([yshift=-0.4cm]shard1.north east);

    \node[body, anchor=east] (shard2) at ([xshift=2.2cm]shard1.east) {
        \\ \\
        $\bullet$ Doc B1 (0.82) \\
        $\bullet$ Doc B2 (0.11) \\
        $\bullet$ Doc B3 (0.05) \\
        $\bullet$ Doc B4 (0.01) \\
        $\bullet$ Doc B5 (0.01)
      };
    \node[header, below=0cm of shard2.north] {\textbf{Shard B}};
    \draw[] ([yshift=-0.4cm]shard2.north west) -- ([yshift=-0.4cm]shard2.north east);

    \node[minimum height=1.971cm] (manyShards) at ([xshift=0.7cm]shard2.east) {\ldots};
    
    \node[body, anchor=east] (shard3) at ([xshift=2.5cm]manyShards.east) {
        \\ \\
        $\bullet$ Doc J1 (0.5) \\
        $\bullet$ Doc J2 (0.39) \\
        $\bullet$ Doc J3 (0.06) \\
        $\bullet$ Doc J4 (0.03) \\
        $\bullet$ Doc j5 (0.02)
      };
    \node[header, below=0cm of shard3.north] {\textbf{Shard J}};
    \draw[] ([yshift=-0.4cm]shard3.north west) -- ([yshift=-0.4cm]shard3.north east);
    
    \node[body, anchor=east] (results) at (3,-2.3) {
        $\bullet$ Doc B1 (0.82) \\
        $\bullet$ Doc A1 (0.6) \\
        $\bullet$ Doc J1 (0.5) \\
        $\bullet$ Doc E1 (0.48) \\
        $\bullet$ Doc G1 (0.45)
      };

    \path[line] (query.south) -- +(0,-0.2) -| (shard1.north);
    \path[line] (query.south) -- +(0,-0.2) -| (shard2.north);
    \path[line] (query.south) -- +(0,-0.2) -| (manyShards.north);
    \path[line] (query.south) -- +(0,-0.2) -| (shard3.north);
    
    \path[line] (shard1.south) -- +(0,-0.2) -| (results.north);
    \path[line] (shard2.south) -- +(0,-0.2) -| (results.north);
    \path[line] (manyShards.south) -- +(0,-0.2) -| (results.north);
    \path[line] (shard3.south) -- +(0,-0.2) -| (results.north);
    
\end{tikzpicture}
    \vspace{0.5em}
    \caption{Confidence-Ensemble using 10 shards (i.e. 10 peer groups). The scores under each shard are post-softmax. The result of the ensemble is the top5 documents with largest post-softmax score.}
    \label{fig:confidenceEnsemble}
\end{figure}
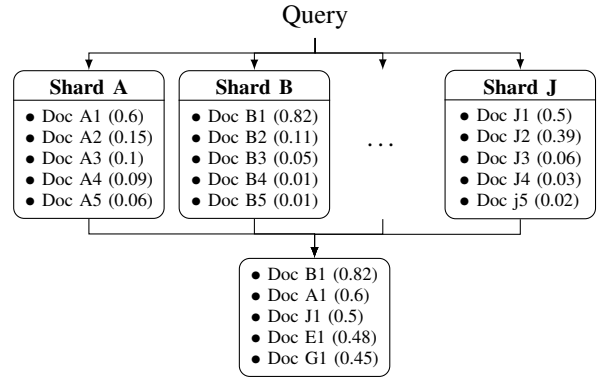

To solve this problem, we propose normalizing the scores using softmax.
Let $D=\{(d_i, s_i) \mid i=1 \leq i \leq 5\}$ be the five result candidates a model generated given some query.
In this set, $d_i$ denotes a generated docid, and $s_i$ represents its logit score.
By taking the softmax of the five scores, we can normalize the confidence scores of the models' suggestions:
\begin{equation}
    D_\text{softmax} = \{ (d_i, \text{softmax}(s_i)) \mid 1 \leq i \leq 5 \} 
\end{equation}
Softmax allows the comparison of results of different models as it scales each of them from zero to one, so model scale variability is nullified. 
Thus, all scores from all models are appended to the same list, and the top-$k$ ones with the highest scores are selected, representing the results of the model for the top-$k$ accuracy metric.
We illustrate this process, dubbed as Confidence-Ensemble, in \cref{fig:confidenceEnsemble}.
Finally, by considering only the suggestions with high confidence scores, we can filter out suggestions from models that have not been trained on related queries.

\section{Performance Evaluation and Experiments}

As the base datasets to our experiments, we use the \textit{Open Resource for the Curation of Answer-Snippets (ORCAS)}~\cite{craswell2020orcas}. 
ORCAS is a comprehensive collection designed to support research in information retrieval, specifically for the development and evaluation of search engines. It contains 1.4 million documents and 20 million query-document pairs, obtained from Bing search histories over the course of a few months. Its data are anonymized and only pertain to English-speaking users within the United States. Despite its relatively low coverage of the entire Internet, it still represents one of the best datasets for analyzing search in the English-speaking world. A few data are shown in \cref{table: Sample for DSI model}.

Throughout this paper the model used was the pre-trained version of the T5-small model (about 60 million parameters) for training/inference. The loss function used in all experiments was the cross-entropy function. We adopt top-$k$ as our main metric of comparison, as explained above. This is because in this dataset we do not have relevance rankings, but only the binary correct/incorrect answer. From our experience, training a T5 model on a dataset of 1000 docids and 40 queries per docid (\qty{40000} data points), takes about 20 hours on a Macbook with M2 Pro, and about 2 hours on an NVIDIA A4000 GPU.

In the following, we are going to evaluate our previously elaborated ideas in four successive experiments. 
First, we are assessing, and furthermore proving, the capability of the T5-small Transformer model to retrieve documents based on unseen queries. 
This evaluation occurs after the model has been trained exclusively on mappings between queries and docids, without ever being exposed to the content of the documents themselves. It is important to mention that this finding stands for web pages similar to that of the ORCAS dataset.
In the second experiment, in an effort to scale up the search engine, we are employing an ensemble of DSI models. 
Only in our third experiment, we are adding more realistic conditions by simulating a P2P system, and we decentralise the training process.
In our fourth and final experiment, we show that the T5-small can also accurately retrieve entire magnet links at low additional cost in terms of accuracy. This is meant to give support to the idea that the model can act as a real search engine.

\subsection{Content-Oblivious Search}
\label{sec:queriesIsAllYouNeed}

We report on a property of LLMs that has never been reported on before. 
Our first experiment shows emergence of an effective search engine merely by \emph{query feeding}. 
The goal of this experiment is to see how many queries are needed to describe a document, such that the model can successfully generalize to new (unseen) queries. 
To this end, we fine-tune our model on a sample of documents and their associated queries. 
With larger samples of documents, there's a higher chance that document may contain similar information and thus be harder to distinguish from one another based solely on queries.
This generally makes it harder for the model to correctly predict the docid given a query, and must always be taken into consideration when assessing the results. 
For that reason, we conduct multiple experiments with sample sizes $N=\{100, 500, 1000\}$. 
Furthermore, we conduct our experiments in a range of $n=1..20$ queries per document.
Naturally, we expect a better performance the more queries are fed in the model.

To start our experiments, we first sample $N$ documents, each with 60 associated unique queries (20/20/20 for training, validation, and testing).
This dataset of query-docid pairs is used as the base throughout the experiments on $n=1..20$.
In every set, the number of queries is equally distributed over all documents.
In each iteration of $n$, we get the 'first' $n$ queries related with each docid, from the train set.
That is, the queries in $n=1$ are also guaranteed to be part of $n=2$, and so on.
A new T5-small model is now fine-tuned on the relationship of every query-docid pair in this subset.
The number of epochs is controlled by a strategy of early stopping,
where the training continues until the accuracy on the validation set has not improved by at least $0.01$ in the last 20 epochs.
Afterwards, the state of the model with the highest accuracy is taken further to the testing phase.
This is generally done to control for overfitting.
Finally, we evaluate the accuracy of the fine-tuned model on the test set. For every docid that the model correctly matches to a query in the test set, we increment a score counter. The score divided by the sum of queries in the test set, reflects the accuracy,
or in other words, the success rate.

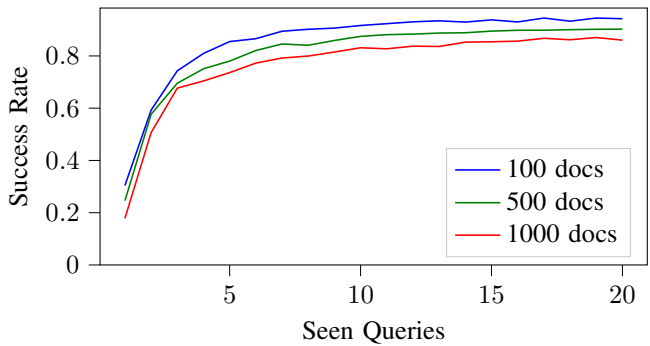
\begin{figure}[h]
    \centering
\begin{tikzpicture}

\definecolor{darkgray176}{RGB}{176,176,176}
\definecolor{green}{RGB}{0,128,0}
\definecolor{lightgray204}{RGB}{204,204,204}

\begin{axis}[
height=5cm,
legend cell align={left},
legend style={
  fill opacity=0.8,
  draw opacity=1,
  text opacity=1,
  at={(0.97,0.03)},
  anchor=south east,
  draw=lightgray204
},
tick align=outside,
tick pos=left,
width=\columnwidth,
x grid style={darkgray176},
xlabel={Seen Queries},
xmin=0.0499999999999999, xmax=20.95,
xtick style={color=black},
y grid style={darkgray176},
ylabel={Success Rate},
ymin=0, ymax=0.983335,
ytick style={color=black}
]
\addplot [semithick, blue]
table {%
1 0.304
2 0.5925
3 0.743
4 0.8095
5 0.855
6 0.866
7 0.8945
8 0.902
9 0.9065
10 0.9165
11 0.9235
12 0.9305
13 0.9345
14 0.9295
15 0.938
16 0.93
17 0.945
18 0.933
19 0.945
20 0.9425
};
\addlegendentry{100 docs}
\addplot [semithick, green]
table {%
1 0.2458
2 0.5766
3 0.6955
4 0.7508
5 0.7803
6 0.8207
7 0.8458
8 0.8408
9 0.8592
10 0.8748
11 0.8814
12 0.8835
13 0.8875
14 0.8887
15 0.895
16 0.8982
17 0.8983
18 0.9005
19 0.902
20 0.9026
};
\addlegendentry{500 docs}
\addplot [semithick, red]
table {%
1 0.1783
2 0.50635
3 0.67685
4 0.7042
5 0.7363
6 0.7725
7 0.792
8 0.7996
9 0.815
10 0.8312
11 0.82735
12 0.8373
13 0.83625
14 0.8528
15 0.8542
16 0.8566
17 0.86745
18 0.86195
19 0.87045
20 0.86035
};
\addlegendentry{1000 docs}
\end{axis}

\end{tikzpicture}%
    \caption{Success rate matching unseen queries to the correct document, based on a number of seen queries trained on.}
    \label{fig:queriesIsAllYouNeed}
\end{figure}

The comprehensive results are depicted in \cref{fig:queriesIsAllYouNeed}.
Generally, we can observe that only very few training queries are needed to answer new queries with remarkable success.
For instance, even with $N=1000$ documents, \textbf{only two queries per document} were needed to answer unseen queries with an accuracy of above \qty{50}{\%}.
After training on 20 queries per document, the success rate has risen to an impressive \qty{86}{\%}, and to \qty{94}{\%} for $N=100$.
All experiments experience a steep logarithmic rise after just the first few queries,
and later seem to stagnate with higher numbers of queries fed. This is probably due to the queries being very similar to one another, so the success rate shrinks with higher values for $N$.

\subsection{Ensemble-DSI for Scalable Search}\label{sec: Ensemble DSI}

As we have seen in the previous experiment, the efficacy of the search engine is subject to the number of documents in the output space.
The purpose of this experiment is to increase the total number of documents in a way that scales.
To this end, we are employing an ensemble of 10 models, 
where each is trained on distinct subsets of the data, called \emph{shards} (i.e. the documents along with all their associated queries).
For each new query, results are solicited from all models, and subsequently aggregated according to the ensemble's design.

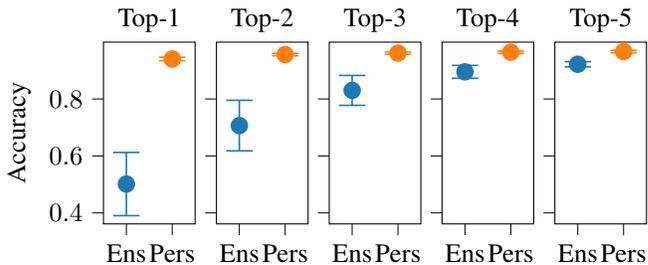
\begin{figure}[hbp]
    \centering
\begin{tikzpicture}

\definecolor{darkgray176}{RGB}{176,176,176}
\definecolor{darkorange25512714}{RGB}{255,127,14}
\definecolor{steelblue31119180}{RGB}{31,119,180}

\begin{groupplot}[
  group style={group size=5 by 1, horizontal sep=8pt},
  title style={yshift=-0.2cm},
]
\nextgroupplot[
height=4cm,
tick align=outside,
tick pos=left,
title={Top-1},
width=2.8cm,
x grid style={darkgray176},
xmin=0.5, xmax=2.5,
xtick style={color=black},
xtick={1,2},
xticklabels={Ens,Pers},
y grid style={darkgray176},
ylabel={Accuracy},
ymin=0.361327613823903, ymax=0.999512039909633,
ytick style={color=black}
]
\path [draw=steelblue31119180, semithick]
(axis cs:1,0.3903359968278)
--(axis cs:1,0.6122640031722);

\addplot [semithick, steelblue31119180, mark=-, mark size=5, mark options={solid}, only marks]
table {%
1 0.3903359968278
};
\addplot [semithick, steelblue31119180, mark=-, mark size=5, mark options={solid}, only marks]
table {%
1 0.6122640031722
};
\path [draw=darkorange25512714, semithick]
(axis cs:2,0.9342)
--(axis cs:2,0.9466);

\addplot [semithick, darkorange25512714, mark=-, mark size=5, mark options={solid}, only marks]
table {%
2 0.9342
};
\addplot [semithick, darkorange25512714, mark=-, mark size=5, mark options={solid}, only marks]
table {%
2 0.9466
};
\addplot [semithick, steelblue31119180, mark=*, mark size=3, mark options={solid}, only marks]
table {%
1 0.5013
};
\addplot [semithick, darkorange25512714, mark=*, mark size=3, mark options={solid}, only marks]
table {%
2 0.9404
};

\nextgroupplot[
height=4cm,
scaled y ticks=manual:{}{\pgfmathparse{#1}},
tick align=outside,
tick pos=left,
title={Top-2},
width=2.8cm,
x grid style={darkgray176},
xmin=0.5, xmax=2.5,
xtick style={color=black},
xtick={1,2},
xticklabels={Ens,Pers},
y grid style={darkgray176},
ymin=0.361327613823903, ymax=0.999512039909633,
ytick style={color=black},
yticklabels={}
]
\path [draw=steelblue31119180, semithick]
(axis cs:1,0.61784505647571)
--(axis cs:1,0.795354943524291);

\addplot [semithick, steelblue31119180, mark=-, mark size=5, mark options={solid}, only marks]
table {%
1 0.61784505647571
};
\addplot [semithick, steelblue31119180, mark=-, mark size=5, mark options={solid}, only marks]
table {%
1 0.795354943524291
};
\path [draw=darkorange25512714, semithick]
(axis cs:2,0.951765994678281)
--(axis cs:2,0.960034005321719);

\addplot [semithick, darkorange25512714, mark=-, mark size=5, mark options={solid}, only marks]
table {%
2 0.951765994678281
};
\addplot [semithick, darkorange25512714, mark=-, mark size=5, mark options={solid}, only marks]
table {%
2 0.960034005321719
};
\addplot [semithick, steelblue31119180, mark=*, mark size=3, mark options={solid}, only marks]
table {%
1 0.7066
};
\addplot [semithick, darkorange25512714, mark=*, mark size=3, mark options={solid}, only marks]
table {%
2 0.9559
};

\nextgroupplot[
height=4cm,
scaled y ticks=manual:{}{\pgfmathparse{#1}},
tick align=outside,
tick pos=left,
title={Top-3},
width=2.8cm,
x grid style={darkgray176},
xmin=0.5, xmax=2.5,
xtick style={color=black},
xtick={1,2},
xticklabels={Ens,Pers},
y grid style={darkgray176},
ymin=0.361327613823903, ymax=0.999512039909633,
ytick style={color=black},
yticklabels={}
]
\path [draw=steelblue31119180, semithick]
(axis cs:1,0.777517778669972)
--(axis cs:1,0.882682221330027);

\addplot [semithick, steelblue31119180, mark=-, mark size=5, mark options={solid}, only marks]
table {%
1 0.777517778669972
};
\addplot [semithick, steelblue31119180, mark=-, mark size=5, mark options={solid}, only marks]
table {%
1 0.882682221330027
};
\path [draw=darkorange25512714, semithick]
(axis cs:2,0.956888765775974)
--(axis cs:2,0.964911234224026);

\addplot [semithick, darkorange25512714, mark=-, mark size=5, mark options={solid}, only marks]
table {%
2 0.956888765775974
};
\addplot [semithick, darkorange25512714, mark=-, mark size=5, mark options={solid}, only marks]
table {%
2 0.964911234224026
};
\addplot [semithick, steelblue31119180, mark=*, mark size=3, mark options={solid}, only marks]
table {%
1 0.8301
};
\addplot [semithick, darkorange25512714, mark=*, mark size=3, mark options={solid}, only marks]
table {%
2 0.9609
};

\nextgroupplot[
height=4cm,
scaled y ticks=manual:{}{\pgfmathparse{#1}},
tick align=outside,
tick pos=left,
title={Top-4},
width=2.8cm,
x grid style={darkgray176},
xmin=0.5, xmax=2.5,
xtick style={color=black},
xtick={1,2},
xticklabels={Ens,Pers},
y grid style={darkgray176},
ymin=0.361327613823903, ymax=0.999512039909633,
ytick style={color=black},
yticklabels={}
]
\path [draw=steelblue31119180, semithick]
(axis cs:1,0.872577886924516)
--(axis cs:1,0.917822113075484);

\addplot [semithick, steelblue31119180, mark=-, mark size=5, mark options={solid}, only marks]
table {%
1 0.872577886924516
};
\addplot [semithick, steelblue31119180, mark=-, mark size=5, mark options={solid}, only marks]
table {%
1 0.917822113075484
};
\path [draw=darkorange25512714, semithick]
(axis cs:2,0.959985738498859)
--(axis cs:2,0.968414261501141);

\addplot [semithick, darkorange25512714, mark=-, mark size=5, mark options={solid}, only marks]
table {%
2 0.959985738498859
};
\addplot [semithick, darkorange25512714, mark=-, mark size=5, mark options={solid}, only marks]
table {%
2 0.968414261501141
};
\addplot [semithick, steelblue31119180, mark=*, mark size=3, mark options={solid}, only marks]
table {%
1 0.8952
};
\addplot [semithick, darkorange25512714, mark=*, mark size=3, mark options={solid}, only marks]
table {%
2 0.9642
};

\nextgroupplot[
height=4cm,
scaled y ticks=manual:{}{\pgfmathparse{#1}},
tick align=outside,
tick pos=left,
title={Top-5},
width=2.8cm,
x grid style={darkgray176},
xmin=0.5, xmax=2.5,
xtick style={color=black},
xtick={1,2},
xticklabels={Ens,Pers},
y grid style={darkgray176},
ymin=0.361327613823903, ymax=0.999512039909633,
ytick style={color=black},
yticklabels={}
]
\path [draw=steelblue31119180, semithick]
(axis cs:1,0.912539330477768)
--(axis cs:1,0.931060669522232);

\addplot [semithick, steelblue31119180, mark=-, mark size=5, mark options={solid}, only marks]
table {%
1 0.912539330477768
};
\addplot [semithick, steelblue31119180, mark=-, mark size=5, mark options={solid}, only marks]
table {%
1 0.931060669522232
};
\path [draw=darkorange25512714, semithick]
(axis cs:2,0.962296343094263)
--(axis cs:2,0.970503656905737);

\addplot [semithick, darkorange25512714, mark=-, mark size=5, mark options={solid}, only marks]
table {%
2 0.962296343094263
};
\addplot [semithick, darkorange25512714, mark=-, mark size=5, mark options={solid}, only marks]
table {%
2 0.970503656905737
};
\addplot [semithick, steelblue31119180, mark=*, mark size=3, mark options={solid}, only marks]
table {%
1 0.9218
};
\addplot [semithick, darkorange25512714, mark=*, mark size=3, mark options={solid}, only marks]
table {%
2 0.9664
};
\end{groupplot}

\end{tikzpicture}
    \vspace{-0.5em}
    \caption{Results of top-$k$ accuracies when inferring from the ensemble (Ens) vs. from only the personal model (Pers), with $k=1..5$.}
    \label{fig:accs_exp2}
\end{figure}

The results of this experiment are shown in \cref{fig:accs_exp2}. The error bars showcase the means and variances calculated over the accuracies of 10 shards, for top-$k$ accuracy with $k=1..5$. The color of the boxplots represents the method, and the y-axis shows the accuracy. 
That is, when asking queries from the test set of any shard, we distinguish between two methods:
\begin{itemize}
    \item \textbf{Ensemble:} Here, the results represent the mean and standard deviation of the accuracy, achieved over all 10 shards. 
    \item \textbf{Personal:} This refers to inference from a single model from within the respective shard (i.e. a model which has trained on queries belonging to the docid we're seeking).
\end{itemize}

It can be seen that, as previously, peers trained well on their data (orange boxplot). In the ensemble we can observe a markedly lower top-1 mean accuracy than in the 'personal model' method. This may be due to other shards containing similar documents (which implicitly have similar queries associated with them). This would lead to the model outputting a sequence with high confidence, even though the expected sequence is not the correct one. For example, the query 'Tesla V3' may render a high confidence from a model which was trained on a document featuring the car, but also on a model which was trained on a document about renowned scientists. In this case there is a high chance that the confidences of both models would be high enough to make the end result a tossup. However, as we take into consideration metrics with a larger $k$ (as in top-$k$), the chance that the right suggestion is among them approaches the chance that the personal model was asked.

We also compare the accuracy of the ensemble vs. the accuracy of a single model trained on all 1000 data in \cref{table: Accs of Simple model vs Ensemble}. Here too, the top-1 accuracy is badly damaged by using the ensemble method. The top-5 accuracy metric interestingly exhibits almost the same result for both the ensemble and the singular T5 method.

\begin{table}
\caption{Comparison of accuracies with 1000 documents, between a single model vs. an ensemble of 10 models.}
\centering
\begin{tabular}{lrr}
\toprule
Model & Top-1 & Top-5 \\
\midrule
Singular & 0.860 & 0.923 \\
Ensemble & 0.501 & 0.922 \\
\bottomrule
\end{tabular}
\label{table: Accs of Simple model vs Ensemble}
\end{table}

In this experiment it does not pay to use the ensemble (both in terms of accuracy and of higher computational cost). However, the high accuracy in the top-5 metric shows that the assembly method of the results of different T5 models works in principle. What would be needed is to find a way to reduce the confusion of models from different shards which most likely affects the metrics presented.

Future research on this topic could attempt to shard data in a semantically meaningful way, possibly by only using queries. This would mean that each shard could deal with a certain aspect of the documents in the dataset, so confusion arising from multiple shards holding similar documents would diminish, thus increasing the accuracy of the ensemble. 

Another research area could be the application of a mixture of experts on the topic of De-DSI, where a `master model' could be utilized to pick which shard to ask the query to. This would increase the scaling capabilities further by not requiring models from all shards to suggest documents. In this case, only models from a few select shards could be made to retrieve answers, reducing the computational complexity required.

\subsection{Decentralised DSI}\label{sec: decentralised Method}

For this experiment we aim to prove the efficacy of our ensemble algorithm in the P2P setting.
To this end, we simulated a network of $N=30$ peers, and divided the data into three shards (i.e. 10 peers per group).
To each shard, we assign \qty{5000} documents. 
We let each peer of that shard randomly sample between 200 and 300 documents from that pool. 
The retrieved set of documents reflects the personal dataset $S$ (comprising query-docid pairs) of a peer. In P2P applications, this would be equivalent to users' recent personal history of search queries and the result they selected.
This way we allow peers to have differently sized datasets,
but also some documents to get sampled by multiple peers,
while others might not got picked at all.
In case of the latter, those documents were discarded from our experiment. 
By these means, we attempt to relax some of the conditions posed in the previous experiment, and add the noise that is encountered in real P2P systems.

Each peer maintains one local T5-small model, and a training batch $B = \{ (q_i, d_j) \mid i, j \in \mathbb{N} \}$. 
The training batch has a fixed size of $|B|=32$, and $(q_i,d_j)$ mark one data point (one query-docid pair).
We train in batches to speed up the training, but also to reduce the noise
that occurs if models train on individual data points.
The training data is collected through gossip within the peer group.
To this end, we perform a simulation in message exchange rounds (in intervals of \qty{0.1}{s}).
In each round, every peer sends one $(q,d)\in S$ to another random peer from its group.
Thereby, each peer receives, on average, one data point per round.
Incoming data points are appended to the local batch.
Furthermore, every new batch is initialized with a sample $S' \subset S$ of size $|S'|=\nint{32/N}$.
The idea behind this strategy is to uniformly distribute the personal dataset throughout the entire training, avoiding overfitting.

This method of peers sending training data to each other doesn't preserve privacy. Our goal in this paper is to showcase the Ensemble-DSI and implement it in a decentralised network. Future work on this topic could look into how to implement the structure of the decentralised algorithm to train the network in a privacy non-invasive manner. One direction worth investigating could be implementing the message exchange along with an onion routing protocol \cite{onionrouting} so that no message travelling through the network could be traced to any one peer. Other measures that could conceivably be used to preserve privacy in decentralized networks are presented in \cite{Hallaji_2024}.

\begin{figure}[htbp]
    \centering
    \resizebox{\columnwidth}{!}{%
        \input{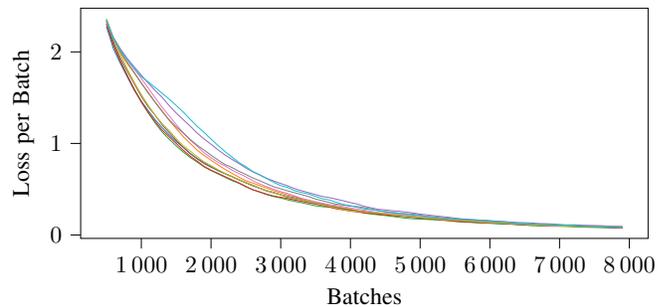}%
    }
    \caption{Rolling mean (with window = 500) of loss per batch for 10 peers within the same shard. Each color represents the loss of one of the peers.}
    \label{fig:10peerloss}
\end{figure}

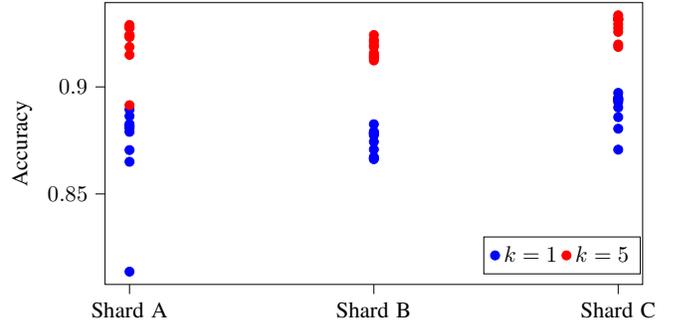
\begin{figure}[htbp]
    \centering
    \resizebox{\columnwidth}{!}{%
\begin{tikzpicture}

\definecolor{darkgray176}{RGB}{176,176,176}

\begin{axis}[
tick align=outside,
tick pos=left,
height=6cm,
x grid style={darkgray176},
xmin=0.9, xmax=3.1,
xtick style={color=black},
xtick={1,2,3},
xticklabels={Shard A,Shard B,Shard C},
y grid style={darkgray176},
ylabel={Accuracy},
ymin=0.807729045648486, ymax=0.93941501708636,
ytick style={color=black},
legend style={at={(0.85,0.17)}, anchor=north, legend columns=-1} 
]
\addplot [draw=blue, fill=blue, mark=*, only marks]
table{%
x  y
1 0.889455782312925
1 0.881741982507289
1 0.870505344995141
1 0.881559766763848
1 0.88265306122449
1 0.886358114674441
1 0.865038872691934
1 0.879008746355685
1 0.880648688046647
1 0.813714771622935
};
\addlegendentry{$k=1$}
\addplot [draw=red, fill=red, mark=*, only marks]
table{%
x  y
1 0.927781827016521
1 0.914965986394558
1 0.891460155490768
1 0.918671039844509
1 0.923408649173955
1 0.924198250728863
1 0.924137512147716
1 0.928935860058309
1 0.923347910592809
1 0.927538872691934
};
\addlegendentry{$k=5$}
\addplot [draw=blue, fill=blue, mark=*, only marks]
table{%
x  y
2 0.878181083732472
2 0.86721680420105
2 0.874372439263662
2 0.866639736857291
2 0.870794621732356
2 0.879162098216862
2 0.88256679554504
2 0.86623578971666
2 0.878411910669975
2 0.877430896185585
};
\addplot [draw=red, fill=red, mark=*, only marks]
table{%
x  y
2 0.912343470483005
2 0.919556812279993
2 0.921114894108142
2 0.914074672514282
2 0.918691211264355
2 0.91580587454556
2 0.922038201858157
2 0.914767153326793
2 0.913382191701772
2 0.924231057764441
};
\addplot [draw=blue, fill=blue, mark=*, only marks]
table{%
x  y
3 0.897205229698932
3 0.892947103274559
3 0.880472592059494
3 0.893846707448723
3 0.885870217104474
3 0.89498620606933
3 0.870696893366919
3 0.894386469953221
3 0.894146575506777
3 0.890368237975291
};
\addplot [draw=red, fill=red, mark=*, only marks]
table{%
x  y
3 0.919815281276238
3 0.93139018831714
3 0.931690056375195
3 0.918675782655632
3 0.927491903562432
3 0.931450161928751
3 0.929291111910759
3 0.931989924433249
3 0.933429291111911
3 0.925692695214106
};
\end{axis}

\end{tikzpicture}%
    }
    \caption{Accuracies on the test set, by shard and beam. Blue dots represent the top-1 accuracy, while red dots show the top-5 accuracy of one peer in the associated shard.}
    \label{fig:accs_per_shard_exp3}
\end{figure}

As the models converged (see \cref{fig:accs_exp2}), we stopped the simulation after \qty{8000} batches have been processed by each peer. Since the training procedure was fundamentally structured around batches, not epochs, we would hope the loss to appear converged at the end of the 8000 batches. We show the loss-per-batch experienced by the all models of one of the shards in Figure \ref{fig:10peerloss}. As can be seen, the training procedure has successfully lowered the loss up to convergence.
In \cref{fig:accs_per_shard_exp3}, we present the accuracies on the test set for all models belonging to each of the three shards (denoted A, B, and C). 
The accuracies found in this experiment are in line with those found in  \cref{sec:queriesIsAllYouNeed}, proving that the decentralised training method was successfully applied, with top-1 averaging \qty{88}{\%}, and top-5 at \qty{92}{\%}.

The testing phase proceeded to sample 3 models from each of the 3 shards (i.e. 9 models in total for each inference), and use the confidence-ensemble (presented in \cref{sec: Ensemble description}) to pick a top-1 and top-5 list for each shard.
Since in this case, 3 models were most likely to output the same docid (because they were trained on the same shard), a simple summing procedure over the post-softmax score of all models was used to calculate the total confidence score for each sequence. 
Only after the sum was calculated, the top-$k$ suggestions of the ensemble were picked.

In \cref{table: decentralised results}, we show the performance of the ensembles. 
Additionally, we conduct an experiment to investigate whether adding more models from the same shard improves accuracy. Specifically, we aim to determine if merging the outcomes of various models trained on identical data leads to enhanced performance.
The label in column ``model pool'' describes whether models exclusively from the same shard were available for sampling, or whether we could sample models from each of the three shards.

\begin{table}[h]
 \caption{Accuracies for our experiment on decentralised DSI training.}
 
 \centering
\begin{center}
\begin{tabularx}{\columnwidth}{crrl}
\toprule
Shard & Top-1 Acc. & Top-5 Acc. & Model Pool \\
\midrule
A & 0.849 & 0.933 & All shards \\
B & 0.850 & 0.932 & All shards \\
C & 0.872 & 0.941 & All shards \\
A & 0.913 & 0.947 & Own shard only \\
B & 0.912 & 0.943 & Own shard only \\
C & 0.922 & 0.950 & Own shard only \\
\bottomrule
\end{tabularx}
\end{center}
\label{table: decentralised results}
\end{table}

It can be seen that the ensemble increases accuracy when used exclusively within the same shard. Because in this ensemble we have three models from each shard, as opposed to a single one, we can see improvements in accuracy of top5 even when compared to the 92\% average shown in \cref{fig:accs_per_shard_exp3}. This is important because it means that the ensemble can be used to improve performance as well, not just to increase the number of available data.
Additionally, the method of pooling models from different shards increases accuracy to levels comparable to using one individual correct model to predict the label. In the case of Top5 it even increases it to levels beyond the performance of the average correct model (found in \cref{fig:accs_exp2}) for all 3 shards.

The amount of computation this ensemble method requires scales linearly with the number of shards in the retrievable dataset (since we need at least one, though as we have seen - more is better, model per shard). This limmits the application of the sharding mechanism. However, the average person uses a search engine 3 to 4 times per day ~\cite{Moz2023}. Assuming one 5-beam query running on a local machine takes about 0.2 seconds (Mac M2 Pro processor), one could distribute the workload of the models in the network. A query written by a peer could be sent to other peers as well so they could 'chime in' with their suggestions. Assuming 20 shards as described above and 4 queries per day for each person, one could send a query to 5 random individuals belonging to each shard. This would lead to a processing time for the activity of online search of about $0.2\cdot4\cdot20\cdot5 = 80$ seconds per day per user, quite a meager amount of processing time per person. 

The overhead for communicating the query and receiving suggestions from peers would be only that required by the transfer of a few bytes, representing the query or docid. Assuming regular internet connection with peers from all shards residing on the same continent, this communication overhead would be placed under 100 ms one way and another 100ms back. Adding the processing time, one search instance would involve a waiting time of around 0.4-0.5 seconds, small enough to make the search convenient.

These results confirm the plausible introduction of sharding data on P2P networks which use an LLM as a main search engine.

\subsection{Decentralised video search}\label{sec: videosearch}

Finally, we demonstrate the generalizability of our method. 
To this end, we added support for content identifiers beyond docids, with a step towards generic URL support. 
This enhancement is particularly significant given the widespread popularity of video services like YouTube and TikTok, which cater to a broad audience. 
BitTorrent provides an open protocol for the decentralised sharing of videos~\cite{bittorrentmeasurements}.
This system identifies files using magnet links that contain a 40-character hexadecimal hash string~\cite{bittorrentBep_0009rst_post}.
We show in this experiment that De-DSI is also capable of generating document identifiers of this type, which are longer than those given by ORCAS' docids (8 characters in length).

\begin{figure}[htbp]
    \centering
    \resizebox{\columnwidth}{!}{%
\begin{tikzpicture}

\definecolor{crimson2143940}{RGB}{214,39,40}
\definecolor{darkorange25512714}{RGB}{255,127,14}
\definecolor{darkslategray38}{RGB}{38,38,38}
\definecolor{forestgreen4416044}{RGB}{44,160,44}
\definecolor{lightgray204}{RGB}{204,204,204}
\definecolor{steelblue31119180}{RGB}{31,119,180}

\begin{axis}[
axis line style={lightgray204},
legend cell align={left},
legend style={at={(0.4,0.42)}, fill opacity=0.8, draw opacity=1, text opacity=1, draw=lightgray204},
tick align=outside,
title={},
x grid style={lightgray204},
xlabel=\textcolor{darkslategray38}{Number of Documents},
xmajorgrids,
xmajorticks=true,
xmin=0.55, xmax=10.45,
xtick style={color=darkslategray38},
xtick={1,5,10},
xticklabels={\qty{1}{K},\qty{5}{K},\qty{10}{K}},
y grid style={lightgray204},
ylabel=\textcolor{darkslategray38}{Accuracy},
ymajorgrids,
ymajorticks=true,
ymin=0.63855, ymax=0.96473,
ytick style={color=darkslategray38},
height=6.5cm
]
\addplot [dotted, line width=1.0pt, darkorange25512714, mark=*, mark size=3, mark options={solid}]
table {%
1 0.9499
5 0.8999
10 0.8605
};
\addlegendentry{docid (top-5)}
\addplot [dotted, line width=1.0pt, crimson2143940, mark=*, mark size=3, mark options={solid}]
table {%
1 0.9473
5 0.8895
10 0.8376
};
\addlegendentry{magnet (top-5)}
\addplot [dotted, line width=1.0pt, steelblue31119180, mark=*, mark size=3, mark options={solid}]
table {%
1 0.9058
5 0.8009
10 0.6953
};
\addlegendentry{docid (top-1)}

\addplot [dotted, line width=1.0pt, forestgreen4416044, mark=*, mark size=3, mark options={solid}]
table {%
1 0.9033
5 0.7773
10 0.6534
};
\addlegendentry{magnet (top-1)}

\end{axis}

\end{tikzpicture}%
    }
    \caption{Comparison of top-1 and top-5 accuracy using the same dataset, with the target string represented as docid and magnet link.}
    \label{fig:docid_vs_magnet}
\end{figure}
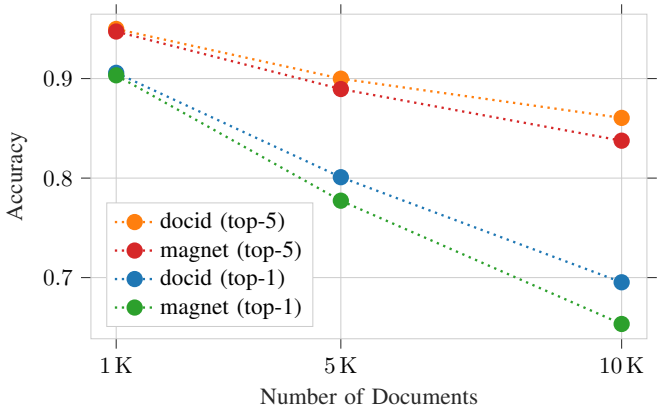

For this experiment, we used a dataset of magnet links based on our prior crawling and dataset efforts from 2003 onwards~\cite{bittorrentmeasurements,100milDHTreplies,p2ptracearchive}.
We merged a magnet dataset with the ORCAS dataset by simply replacing docids with URLs. It needs to be mentioned that the URL's characters are not semantically relevant to the document they refer to, similar to the ORCAS docids in this sense.
We expected that generating URLs would be error-prone as there are more tokens which have to be generated correctly in order to identify a document. If one of the 40 characters is mismatched, we count that as a failure.

The experiment follows the design in \cref{sec:queriesIsAllYouNeed}. We tested with \qty{1000}, \qty{5000}, and \qty{10000} documents, with their ID encoded either in the default ORCAS way, or with an assigned magnet link. The documents have been chosen so they have at least 40 queries associated with them, 20 in the training set, 10 in validation, and 10 in the test set. The results shown in \cref{fig:docid_vs_magnet} are for top-1 and top-5 accuracies on the test set.
When the dataset is relatively small, the accuracies are the same for both top-1 and top-5. As more data appears in the dataset, we can see a divergence in the accuracies posted in both metrics. 
We hypothesize that the limited number of weights in our model efficiently captures URL patterns in scenarios with sparse data. However, as the data complexity increases, this constraint appears to hinder the model's ability to accurately recall the exact sequence of tokens in each URLs.
This is merely a guess, and we intend to investigate this further in future work.
However, the observed discrepancy in accuracy levels remains marginal, amounting to merely a few percentage points across a corpus of \qty{10}{K} documents.

It is important to mention that in a peer-to-peer network, using magnet links to refer to documents/videos is going to exclude the possibility to use semantically meaningful docids. This means that algorithm improvements would need to come from other places.

These preliminary results indicate that intermediaries such as video-sharing platforms can be decentralised. Our experimental work indicates that many entertainment platforms, e-commerce marketplaces, and financial intermediaries \emph{could} be replaced with decentralised generative AI and various tools for decentralisation~\cite{abbas2009gossip,bambacht2022web3,de2022decentralizing}. 

\section{Related Work}

\begin{table*}[ht]
 \caption{Technological progress of AI for information retrieval}
\begin{center}
\begin{tabularx}{\linewidth}{lllccccc}
\toprule
Date & Title & Inventor & Generative AI & Lifelong Learning & Decentralised &  Web Scale \\
\midrule
Sep 2011 & SGD~\cite{Ormándigossiplearning} & Szeged Univ. & - & - & \checkmark & -  \\
Feb 2022 & DSI~\cite{tay2022transformer} & Google & \checkmark & - & - & -  \\
Mar 2022 & DynamicRetriever~\cite{zhou2023dynamicretriever} & Renmin Univ. & \checkmark & - & \emph{partial} & -  \\
Apr 2022 & SEAL~\cite{bevilacqua2022autoregressive} & Meta & \checkmark & - & - & -  \\
Jun 2022 & NCI~\cite{wang2022neural} & Microsoft & \checkmark & - & - & -  \\
Dec 2022 & DSI++~\cite{mehta2022dsi++} & Google & \checkmark & \checkmark & - & -  \\
Apr 2023 & GenRet~\cite{sun2024learning} & Baidu & \checkmark & - & - & - \\
\textbf{Apr 2024} & \textbf{De-DSI} & \textbf{Our team} & \CheckmarkBold & - & \CheckmarkBold  & - \\
\bottomrule
\end{tabularx}
\end{center}
\label{table:technologicalProgress}
\end{table*}

The potential of scaling up model-based retrievers by employing a distributed model has also been addressed in another study.
Zhou et al.~\cite{zhou2023dynamicretriever} proposed \textit{DynamicRetriever},
a model which showed improved accuracy over even the most advanced variant of DSI (DSI with semantically structured docids).
In their study, the authors randomly partitioned a collection of \qty{3.2} million documents into 32 distinct subsets, each containing \qty{100000} documents.
Each subset functioned as the training set for a model.
That is, 32 individual models were trained on different datasets, respectively.
This allowed the models to be much smaller,
as fewer documents had to be memorized on an individual basis.
In the distributed setting,
groups of peers would be assigned different models,
thus training on different subsets of the data.
At retrieval, the query is sent to each group,
and from each model a list of the top 100 documents, with their relevance scores, is retrieved.
These items (\qty{3200} in total) are merged into a final ranking list.
Their experiments yielded a sharp decline in accuracy over the non-distributed setting.
The authors concluded that this is due to the inconsistent scale of scores learned by the independently trained models. Our model used softmax to aggregate the results, thus comparing them on the same scale.

The advent of DSI has motivated other researchers to develop techniques and advancements that would yield better retrieval accuracy.
It has been shown in the architectures for DSI-QG~\cite{zhuang2022bridging} and NCI~\cite{wang2022neural}, for instance, that the generation and feeding of artificial queries on the basis of the documents' contents has the capability to significantly improve retrieval performance \cite{zhuang2022bridging,wang2022neural}.
Furthermore, Meta proposed SEAL~\cite{bevilacqua2022autoregressive}, a system that uses n-grams from documents as docids,
effectively improving efficacy.
The performance of SEAL has been topped in GenRet~\cite{sun2024learning},
where an autoencoder is trained to tokenize documents into semantic docids.

While all those efforts have proved to be effective,
they rely on the knowledge of document contents,
which usually is not given in decentralised applications.
In addition to techniques that exploit this knowledge, however,
the authors of NCI~\cite{wang2022neural} also proposed a novel prefix-aware weight-adaptive (PAWA) decoder, 
as well as an updated regularization loss function.
Both have shown positive effects on the search engine's performance.
As those experiments have been conducted on the basis of semantic docids,
it is not clear what the effects of the PAWA encoder or the altered loss function would be on De-DSI.

Finally, DSI++~\cite{mehta2022dsi++} addresses lifelong learning in the context of DSI.
To this end, they propose two solutions.
Firstly, they leverage \textit{Sharpness-Aware Minimization (SAM)} to steer the model towards flatter loss basins, enhancing stability and reducing the propensity for catastrophic forgetting.
Secondly, they employ a generative memory designed to produce pseudo-queries based on previously indexed documents.
These pseudo-queries are then utilized for rehearsal, further bolstering the model's ability to retain and recall information over extended periods.

We have summarised these developments in \cref{table:technologicalProgress}.
As can be seen, De-DSI is the first work to take a step into decentralizing generative AI for retrieval at web scale.

\section{Conclusion}

With De-DSI, we merged two fields and simultaneously brought them a small step closer to user trials and broad societal usage. 

First, research into DSI represents a significant improvement in efficiency and efficacy compared to the classical index-retrieve-rerank architecture of information retrieval.
By removing the need for document term indexing, and training merely on query-docid pairs, we eased the process further. Our key finding is that the mere provision of queries is sufficient to turn an open source Transformer into a public search engine. Additionally, we find that magnet links can directly be retrieved by the DSI method, with minimal impact on accuracy given a relatively small dataset.

Secondly, our De-DSI ensemble model shows self-scaling properties in our experiments. Although the computational complexity of the ensemble increases linearly with the number of shards in the retrievable dataset, we envision a distribution of the workload within the network to address this issue \cite{zeilemaker2013building}. 
Each part of the global network could specialise in a certain flavour of content and build-up of a stable community. This stable community in turn enables strong security, for instance, with self-sovereign identities~\cite{stokkink2018deployment}
and state-of-the-art Sybil-tolerant trust frameworks~\cite{nasrulin2022meritrank}.

\section*{Acknowledgment}

This work was funded by Dutch national NWO/TKI science grant BLOCK.2019.004.


\printbibliography

\end{document}